\documentclass[preprint,superscriptaddress,aps,longbibliography]{revtex4-2}
\usepackage[utf8]{inputenc}
\usepackage[T1]{fontenc}
\usepackage{amsmath}
\usepackage{amssymb}
\usepackage{graphicx}
\usepackage[capitalize, nameinlink]{cleveref}
\usepackage{subfigure}
\graphicspath{{figures/}}
\usepackage{adjustbox}
\usepackage{booktabs}
\usepackage{multirow}

\usepackage{color}
\usepackage{rotating}
\usepackage{ulem}

\begin{document}

\title{Accelerating magnonic simulations with the pseudospectral Landau-Lifshitz equation}

\author{A. Roxburgh}
\affiliation{Center for Magnetism and Magnetic Nanostructures, University of Colorado Colorado Springs, Colorado Springs, CO, USA}
\author{M. Copus}
\affiliation{Center for Magnetism and Magnetic Nanostructures, University of Colorado Colorado Springs, Colorado Springs, CO, USA}
\author{E. Iacocca}
\email{eiacocca@uccs.edu}
\affiliation{Center for Magnetism and Magnetic Nanostructures, University of Colorado Colorado Springs, Colorado Springs, CO, USA}

\date{\today}

\begin{abstract}
The pseudospectral Landau-Lifshitz (PS-LL) model can describe atomic-scale magnetic exchange interactions within a continuum framework. This is achieved by employing a convolution kernel that models the nonlocal interaction in a grid-independent manner. Even though the PS-LL was originally introduced to address atomic exchange, any nonlocal kernel can be modeled. In the field of magnonics, the dipole field is fundamental to describe the dispersion relation of magnons, the quasiparticle representation of angular momentum. Because dipole-dipole interactions are long-range, numerical approaches typically rely on convolutions. Here, we demonstrate that the PS-LL model can be used to perform magnonic simulations with a single convolution kernel derived from analytical solutions. We demonstrate a twofold increase in computational speed compared with the full dipole calculation. This approach is valid insofar as the excitations are linear, which is typically the case for magnons. Our results have the potential to accelerate magnonic research, particularly for the inverse design method, where several simulations must be performed to achieve the desired outcome.
\end{abstract}

\maketitle

\section*{Introduction}

The field of magnonics~\cite{Neusser2009,Lenk2011,Demokritov2013} studies the control of the quanta of angular momentum, magnons, for practical applications~\cite{Chumak2015,Chumak2022,Flebus2024}. The main interest is to utilise angular momentum as an information carrier in contrast to electronics, which utilise the electron's charge. This approach leads to two benefits: a reduction in Joule heating and the ability to naturally distinguish between two spin states. Despite demonstrating analogies to electronic circuits~\cite{Chumak2013,Chumak2015}, one of the main benefits in magnonics is the possibility to engineer a magnonic crystal~\cite{Krawczyk2014} and thus reconfigure the magnons' properties~\cite{Grundler2015,Karenowska2012,Wagner2016,Wang2017,Szulc2022}. Reconfigurable magnonic band structures have been demonstrated in several types of magnonic crystals~\cite{Tacchi2011,Zivieri2011,Tacchi2012,Kumar2014,Graczyk2018,Gubbiotti2018,Negrello2022,Micaletti2025,Micaletti2025b} as well as artificial spin ices~\cite{Iacocca2016,Iacocca2017c,Lendinez2019,Iacocca2020,Gliga2020,Gartside2021,Lendinez2023,Micaletti2023,Alatteili2023,Alatteili2024}, including applications in reservoir computing~\cite{Gartside2022,Dion2024,Lee2024,Stenning2024}. The advent of thermal nanolithography to spatially modify the magnetic parameters of engineered bilayers~\cite{Albisetti2016,Levati2023} has also led to the investigation of nano-magnonic crystals~\cite{Roxburgh2024,Roxburgh2025}. Reconfiguration is key to the inverse design approach for magnonic devices~\cite{Wang2021,Papp2021,Zenbaa2025} where defects in a magnetic material are determined by machine learning to achieve desired functionalities.

Despite the quasiparticle interpretation of magnons, their manifestation is delocalised, establishing waves throughout magnetic materials. These waves have well-defined dispersion relations that are the result of nonlocal interactions. The simplest representation of a magnon takes into account the Heisenberg exchange interaction between nearest neighbour, localised magnetic moments~\cite{White2007}. This leads to an isotropic dispersion relation that is physically valid for short-wavelength magnons, approximately below $100$~nm. However, typical magnons excited with microwave antennas have wavelengths on the order of micrometres. For such magnons, the Heisenberg exchange interaction between atomic sites is negligible, and instead the dipole interaction dominates. These magnetostatic, or dipolar, magnons exhibit an anisotropic dispersion relation that depends on the wave's propagation direction and the direction of applied fields~\cite{Nikin1974,Stancil2009}.

Analytically, the dispersion relation of magnetostatic waves can be readily obtained for the three limiting cases found in an easy-plane material~\cite{Stancil2009}: forward volume waves (FVW) when the wavevector is in-plane and the magnetic field is normal to the plane, backward volume waves (BVW) when the field is in-plane and the wavevector is parallel to it, and surface waves (SW), or Damon-Eshbach waves, when the field is in-plane and the wavevector is perpendicular to it. Solutions require numerical computations in the full two-dimensional (2D) case as well as in cases approaching the exchange-dominated limit~\cite{Kalinikos1986,Harms2022} or when coupling to elastic modes is considered~\cite{Camley1978,Camley1979}.

In simulations, the dispersion relations of magnetostatic modes~\cite{Venkat2013} can be directly obtained by including dipole interactions. A common method in finite-difference approaches is to use the Newell tensor~\cite{Newell1993} which is geometric in origin and applied as a convolution to the magnetisation. While the tensors need only be computed once due to their geometric origin, the convolution is typically performed in Fourier space, requiring six fast Fourier transforms per time step of the integration algorithm. These computations constitute one of the main bottlenecks in micromagnetic simulations, limiting the achievable throughput. In approaches such as inverse design, such limitations can require significant computational time.

Here, we demonstrate an alternative approach whereby analytically derived dispersion relations are used directly to simulate magnons and their interactions. This approach is based on the pseudospectral Landau-Lifshitz (PS-LL) method~\cite{Rockwell2024} in which the dispersion relation serves as a convolution kernel to model nonlocal interactions. In this case, we implement a ``dipole-exchange'' kernel that describes BVWs and SWs in a 2D simulation domain, as well as their transition to an exchange-dominated regime. We demonstrate that our approach captures the correct physics by comparing results with micromagnetic simulations using the magnon dispersion relation and full computation of the nonlocal dipole field. By computing fewer fast Fourier transforms per time-integration step, our code achieves a speedup of a factor 2 for a large number of cells. One limitation is that use of the dipole-exchange kernel is appropriate only insofar as the dynamics remain within the linear regime. In other words, physically meaningful solitons cannot be nucleated with this approach. Use of the dipole-exchange kernel can nevertheless help accelerate magnonic research and train machine learning algorithms for inverse design magnonics.

\section{PS-LL Method}

The PS-LL method~\cite{Rockwell2024} represents magnetisation dynamics using a modified Landau-Lifshitz equation
\begin{equation}
\label{eq:LL}
\frac{\partial}{\partial t}\mathbf{m} = -\mathbf{m}\times\mathbf{H}_\mathrm{eff} - \alpha\mathbf{m}\times\left(\mathbf{m}\times\mathbf{H}_\mathrm{eff}\right)
\end{equation}
where $\mathbf{m}$ denotes the magnetisation vector normalised by the saturation magnetisation, $M_s$. The first term represents the conservative motion while the second term, scaled by the Gilbert damping constant $\alpha$, accounts for energy dissipation to the lattice. The effective field field $\mathbf{H}_\mathrm{eff}$ is given by
\begin{equation}
    \label{eq:Heff}
    \mathbf{H}_\mathrm{eff} = \gamma \mu_0 M_s \mathbf{h}_l - \mathcal{F}^{-1}\left\{\kappa(\mathbf{k})\mathbf{\hat{m}}\right\},
\end{equation}
where $\gamma$ is the gyromagnetic ratio and $\mu_0$ is the vacuum permeability. The field is separated into local and nonlocal contributions. In this work, the local contribution includes a normalised external magnetic field of magnitude $H_0$ oriented in the plane of the film and the demagnetisation field. The nonlocal contributions are modelled by a convolution kernel $\kappa(\mathbf{k})$, $\mathbf{\hat{m}}$ is the Fourier transform of the magnetisation, and $\mathcal{F}^{-1}$ denotes the inverse Fourier transform. In previous works, the convolution kernel has been set to the magnon dispersion relation in one dimension (1D)~\cite{Rockwell2024} or 2D~\cite{Foglia2024} as well as including the symmetry-breaking Dzyaloshinskii-Moriya interaction~\cite{Copus2025}. Here, we are primarily interested in dipole-exchange waves, and we therefore derive a kernel that represents their 2D dispersion relation. It is worth noting that the dipole contribution cannot simply be added to the dispersion relation, as detailed below.

The dispersion relation for dipole-exchange waves was derived by Kalinikos and Slavin~\cite{Kalinikos1986}. For a thin film, it is sufficient to consider the lowest-energy mode in infinitely extended media
 \begin{equation}
       \label{eq:Slavin0}
    \omega_\mathrm{DE}^2(\mathbf{k}) = [\omega_H+\omega(\mathbf{k})][\omega_H+\omega(\mathbf{k})+\omega_MF(\mathbf{k})],
\end{equation}
where $\omega_H=\mu_0|\gamma|H_0$ is the Larmor frequency associated with the in-plane applied magnetic field of magnitude $H_0$, and $\omega_M=\mu_0|\gamma|M_s$ is the frequency associated with the saturation magnetisation $M_s$. For our 2D system, the function $\omega(\mathbf{k})$ is set to be the 2D magnon dispersion relation given by
\begin{equation}
    \label{eq:omega2D}
    \omega(\mathbf{k}) = -|\gamma|\mu_0M_s2\left(\frac{\lambda_\mathrm{ex}}{a}\right)^2\left(2-\cos{(ak_x)}-\cos{(ak_y)}\right),
\end{equation}
where $\mathbf{k}=(k_x,k_y)$ is the in-plane wavevector, $\lambda_\mathrm{ex}$ is the exchange length, and $a$ is the lattice constant. The function $F(\mathbf{k})$ in Eq.~\eqref{eq:Slavin0} accounts for dipolar interactions and is given by
 \begin{equation}
    \label{eq"Fnn}
    F(\mathbf{k})=1-P(\mathbf{k})\cos^2{(\varphi)}+\omega_M\frac{P(\mathbf{k})(1-P(\mathbf{k}))\sin^2{(\varphi)}}{[\omega_H+\omega(\mathbf{k})]},
\end{equation}
 where $\varphi$ describes the in-plane angle with respect to the $x$-axis. The function $P(\mathbf{k})$ depends on the pinning conditions at the film surfaces. We consider the case of totally unpinned boundaries, which leads to
 \begin{equation}
    \label{eq:Poo}
    P(\mathbf{k})=1-\frac{1}{|\mathbf{k}|d}\left[1-e^{-|\mathbf{k}|d}\right],
\end{equation}
where $d$ is the film thickness and $|\mathbf{k}|$ is the magnitude of the in-plane wavevector.

\begin{figure}
\includegraphics[width=3.3in]{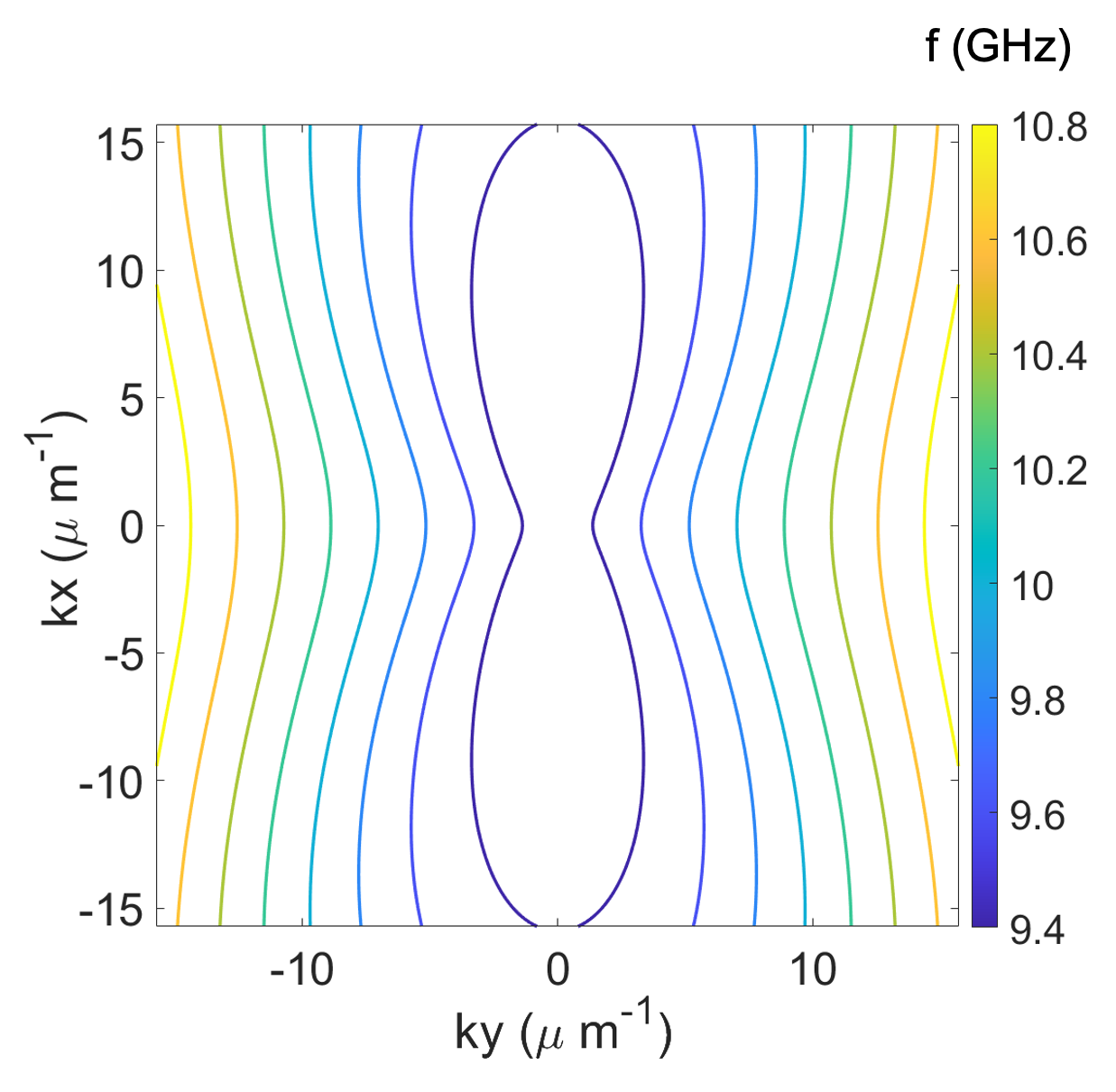}
\centering
\caption{\label{PeanutPS-LL} \textbf{Contour of the dispersion relation for in-plane magnons}. The BVW propagates along the $k_x$ direction while the SW propagates along the $k_y$ direction. The contour displays the distinctive anisotropic shape of the dispersion relation.}
\end{figure}

The solution of Eq.~\eqref{eq:Slavin0} is plotted as a contour map in Fig.~\ref{PeanutPS-LL} for permalloy material parameters, i.e., $M_s=790$~kA~m$^{-1}$, exchange constant $A=10$~pJ~m$^{-1}$ leading to $\lambda_\mathrm{ex}=5$~nm, a lattice constant $a=0.4$~nm, and a film thickness $d=10$~nm. An applied field $H_0=79$~kA~m$^{-1}$ is applied along the $x$-axis leading to BVWs along $k_x$ direction and SWs along the $k_y$ direction. Because of the exchange contribution, both BVWs and SWs blueshift as $|\mathbf{k}|$ increases, leading to the well-known minimum in the magnon dispersion relation where magnon Bose-Einstein condensates can form~\cite{Demokritov2001,Bozhko2016}.

To incorporate Eq.~\eqref{eq:Slavin0} into the PS-LL model, we derive a suitable convolution kernel with an in-plane external field. In this case, the demagnetising field in Eq.~\eqref{eq:Heff} is considered to be $-m_z\hat{z}$, which is an approximation for thin films. This implies that the static contribution of the demagnetising field is local, while the nonlocal contribution is only considered in the convolution kernel. Taking the Fourier transform of the conservative term in Eq.~\eqref{eq:LL} and solving the resulting eigenvalue problem yields the wave frequency (see Methods for details)
\begin{equation}
    \label{eq:omega_nl}
    \omega^2_\mathrm{DE}(\mathbf{k})=\left[\omega_H+\kappa(\mathbf{k})\right]\left[\omega_H+\omega_M+\kappa(\mathbf{k})\right].
\end{equation}

From this, the ``dipole-exchange'' kernel can be written as the positive root of the resulting quadratic equation
\begin{equation}
    \label{eq:kernel}
    \kappa(\mathbf{k}) = -\omega_H-\frac{\omega_M}{2}+\frac{\sqrt{\omega_M^2+4\omega^2_\mathrm{DE}(\mathbf{k})}}{2},
\end{equation}
where the positive sign is chosen to ensure that $\kappa(\mathbf{k}) \rightarrow 0$ as $\mathbf{k}\rightarrow0$. This condition is essential because the ferromagnetic resonance is the result of the local field contributions in Eq.~\eqref{eq:Heff}. It must also be noted that this kernel is only valid for in-plane external fields, for which the uniaxial demagnetising field approximation is appropriate.

\begin{figure}
\includegraphics[width=6.6in]{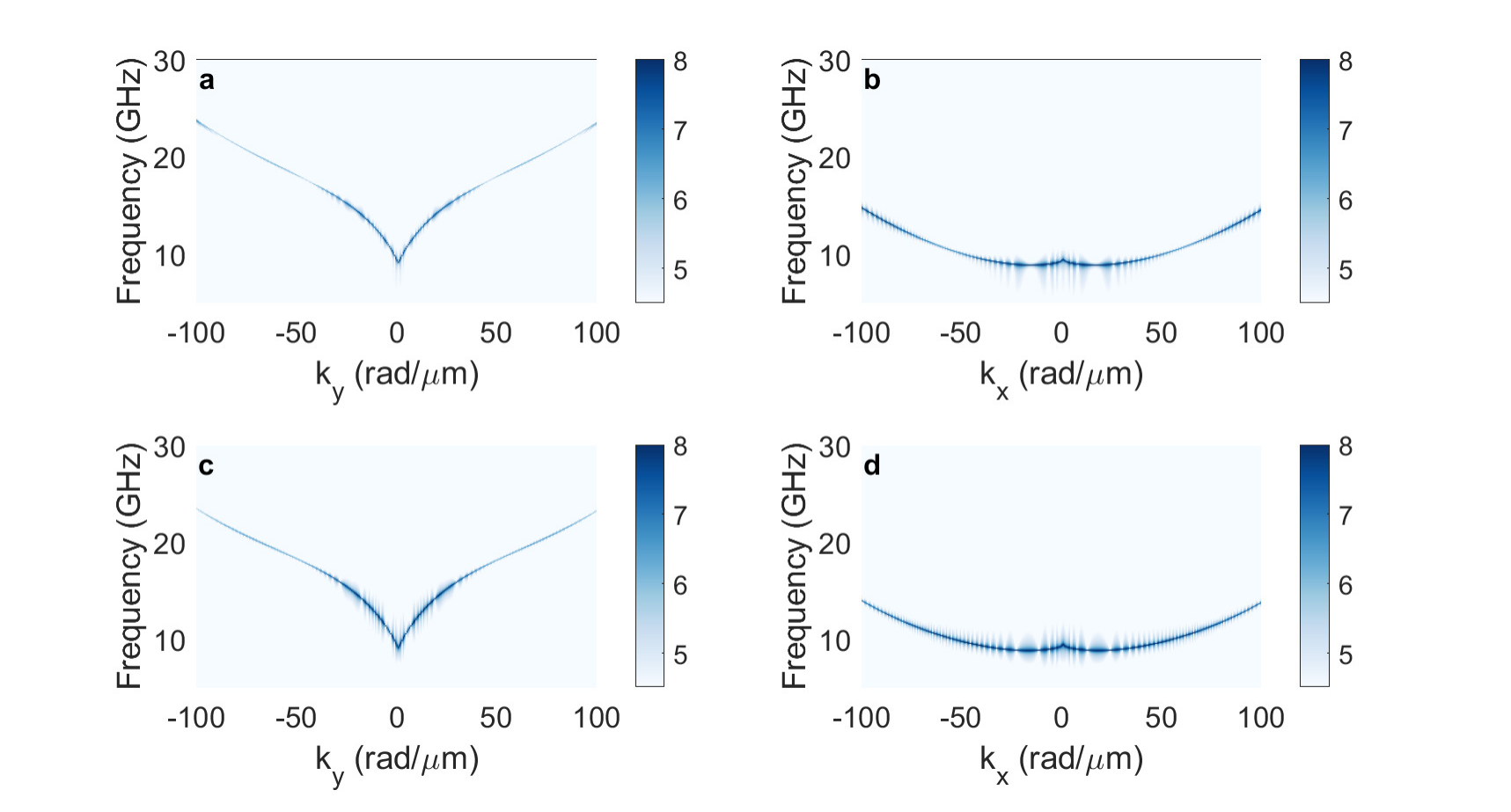}
\centering
\caption{\label{fig:SlavinNewell3D} \textbf{Dispersion relations.}  Line cuts of the dispersion relation using the dipole-exchange kernel for the \textbf{a} SW when $k_x=0$ and \textbf{b} BVW when $k_y=0$. The equivalent simulations using the magnon kernel and full computation of the nonlocal dipole is shown in panels \textbf{c} and \textbf{d}, respectively. The colour scheme is found in Ref.~\cite{colorscheme}.}
\end{figure}

We verify the implementation of the dipole-exchange kernel by numerically computing the 2D dispersion relation. For this, we set a simulation domain of $1$~$\mu$m~$\times1$~$\mu$m discretised in square cells of $20$~nm in lateral size. With this, we obtain a maximum wavevector of approximately $157$~rad/$\mu$m and a wavevector resolution of $3.14$~rad/$\mu$m. The dispersion relation is obtained by initialising the system in a uniform $m_x$ component perturbed with a narrow Gaussian pulse in the $m_z$ component. We implement periodic boundary conditions for this simulation, as detailed in the Methods. A three-dimensional (3D) surface of the resulting dispersion is obtained. Line cuts along the $k_y$ and $k_x$ axes are shown in Fig.~\ref{fig:SlavinNewell3D}\textbf{a} and \textbf{b}, respectively, for SWs and BVWs. These curves are in quantitative agreement with the analytical dispersion relation, and confirm the correct implementation of the dipole-exchange kernel of Eq.~\eqref{eq:kernel}. 

To further validate this approach, we compare results using the dipole-exchange kernel with numerical simulations using the 2D magnon dispersion relation, Eq.~\eqref{eq:omega2D}, as the kernel and the demagnetising field computed from the full non-local dipole field via Newell tensors~\cite{Newell1993}. Line cuts along the $k_x$ and $k_y$ directions are shown in Fig.~\ref{fig:SlavinNewell3D}\textbf{c} and \textbf{d}, respectively. We find excellent agreement between the two approaches, providing additional verification of the accuracy of the dipole-exchange kernel.

A fundamental limitation of our approach is that the dipole-exchange kernel only describes magnons, which are linear excitations. While the PS-LL model of Eq.~\eqref{eq:LL} is nonlinear, the choice of a kernel describing linear excitations will incorrectly describe nonlinear solutions such as solitons~\cite{Kosevich1990}, e.g., magnetic droplets~\cite{Hoefer2010,Mohseni2013}, magnetic skyrmions~\cite{Bogdanov2001,Fert2013}, and domain patterns~\cite{Hubert2009}.

From a computational point of view, the main advantage of using the dipole-exchange kernel is the reduction in the number of fast Fourier transform (FFT) computations required per integration step. Using the dipole-exchange kernel entails six FFTs per step, three to transform the magnetisation components to Fourier space and three to perform the inverse Fourier transform after application of the kernel. In contrast, using the Newell tensor in our implementation requires twelve FFTs per step, the same six as for the dipole-exchange kernel plus an additional six FFT computations, three to transform the magnetisation with boundary conditions, and three inverse transforms after applying the tensor, which cannot be simply combined with the exchange kernel because of different matrix sizes and boundaries. To obtain a benchmark for performance, we computed the dispersion relation for varying domain sizes using a laptop CPU without any parallelization.

\begin{figure}
\includegraphics[width=3.3in]{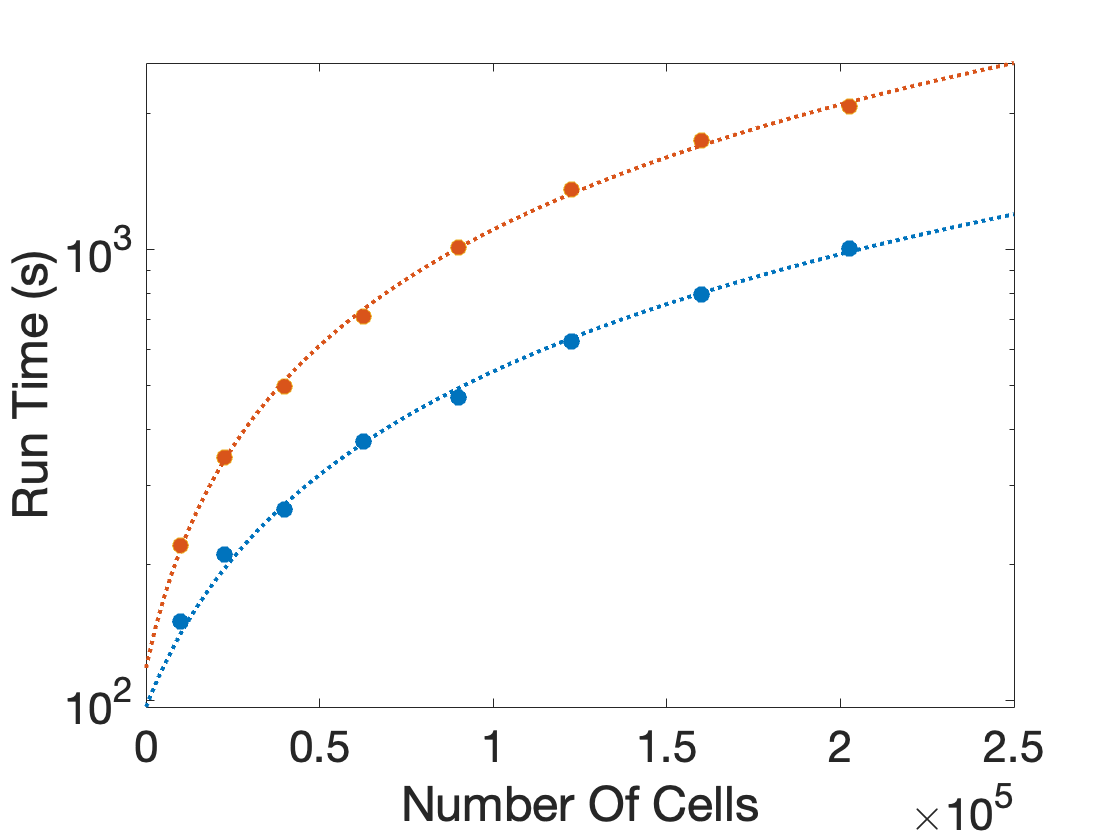}
\centering
\caption{\label{fig:TimePS-LL} \textbf{Computation time comparison.} Computation time using the dipole-exchange kernel (blue symbols and dotted curve) and the exchange kernel with nonlocal dipole field computation (red symbols and dashed curve) as a function of number of cells.}
\end{figure}

The the computation time on a logarithmic scale for each case as a function of the total number of cells is displayed in Fig.~\ref{fig:TimePS-LL}. For domains exceeding $10^5$ cells, we observe a twofold reduction in computation time using the dipole-exchange kernel compared to the use of a full nonlocal dipole field. Parallelization can further increase this difference and reduce the computation time dramatically.

\section*{Simulation examples}

For this method to be beneficial in magnonics research, externally forced magnons and their physical behaviour must be accurately represented. We present three examples of magnonic simulations that are possible to solve with our model.

\subsection*{Wave excitation}

We begin by replicating the traditional case of magnons excited along a nanowire. We consider an elongated domain of $10$~$\mu$m$\times 200$~nm. A static effective field of 79~kA~m$^{-1}$ is applied in-plane. The waves are excited by an oscillating $11$~GHz external field defined in the centre of the sample over a 200~nm-wide region, applied in-plane and normal to the static field. The oscillating field's frequency is chosen above the FMR frequency of 9.29~GHz so that the excited BVWs and SWs have distinct wavenumbers. In these simulations, we implement periodic boundary conditions.

\begin{figure}
\includegraphics[width=6.6in]{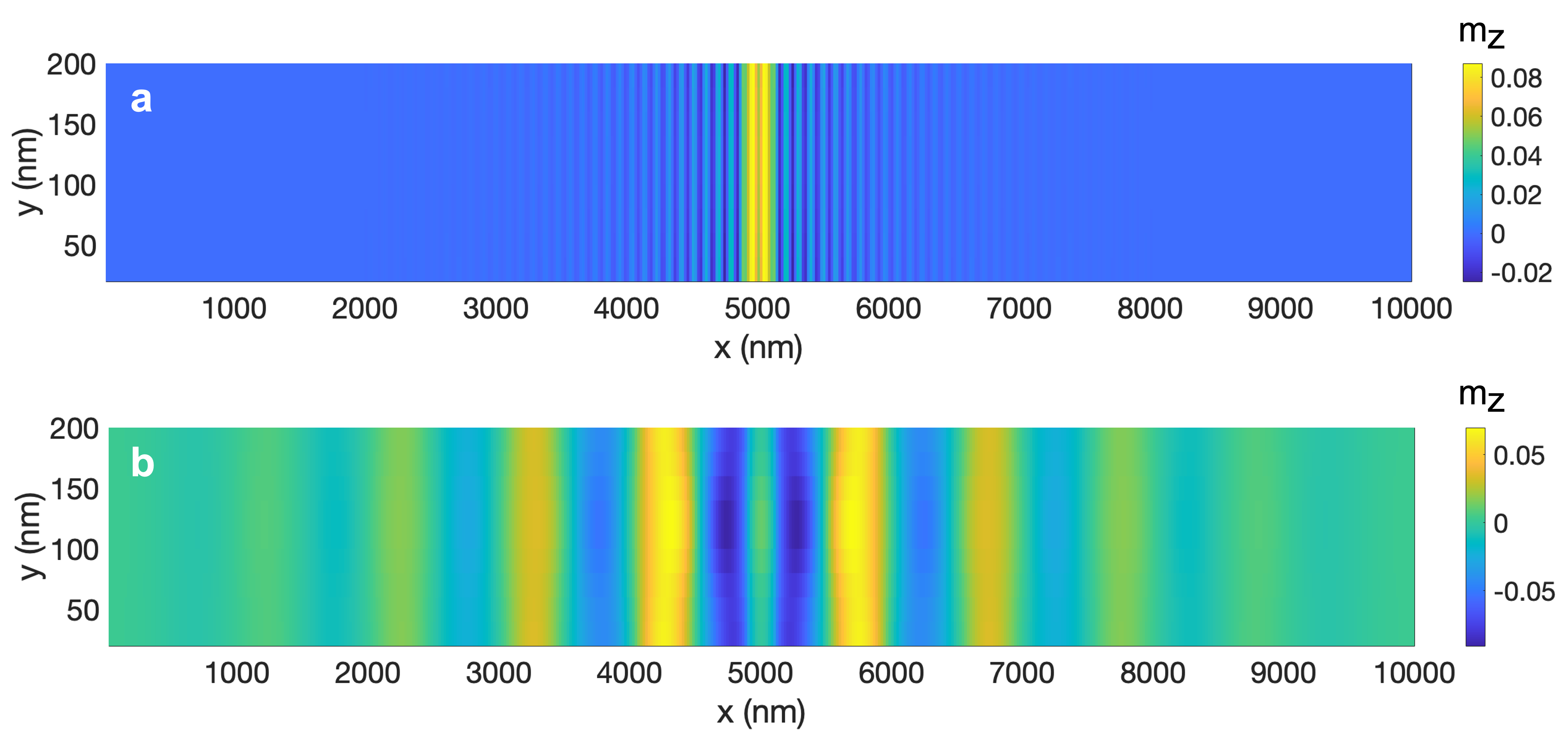}
\centering
\caption{\label{WavesInNanoWire} \textbf{Wave excitation in a nanowire.} Two films displaying waves in a nanowire in the \textbf{a} BVWM configuration with the oscillating field directed in plane and parallel to wave propagation, and the \textbf{b} SWM configuration, with the oscillating field directed in plane and perpendicular to wave propagation.}
\end{figure}

To excite a BVW, we apply the static field along the $x$-axis and the oscillating field along the $y$-axis. A snapshot of the resulting wave is shown in Fig.~\ref{WavesInNanoWire}\textbf{a} by the $m_z$ magnetization component. The wavenumber is large and estimated by Fourier analysis to $65.9\pm 1.3$~rad~$\mu$m$^{-1}$, in excellent agreement with the expected $64.7$~rad~$\mu$m$^{-1}$. To excite a SW, we apply the static field along the $y$-axis and the oscillating field along the $x$-axis. A snapshot of the $m_z$ component is shown in Fig.~\ref{WavesInNanoWire}\textbf{b}. In this case, the wavenumber is small and estimated to $6.4\pm 1.3$~rad~$\mu$m$^{-1}$, also in good agreement with the expected $5.9$~rad~$\mu$m$^{-1}$.

These results show that our model can excite simple waves which are relevant for nanoscale microwave signal processing~\cite{Verba2016} and spintronic devices~\cite{Gubbiotti2019}.

\subsection*{Caustic waves}

Caustic waves are spin wave beams that radiate from a point-source excitation along preferred directions~\cite{Bible2017}. This phenomenon occurs at relatively high frequencies, where the dispersion relation approaches the exchange-dominated regime and the traditional ``peanut'' shape of magnetostatic waves, see e.g. Fig.~\ref{fig:SlavinNewell3D}, become nearly elliptical. This means, for surface waves at the edge of the isofrequency curve, the dispersion relation is effectively flat in $k$-space, suppressing scattering and focusing the energy into narrow beams.

We generated caustic beams with the dipole-exchange kernel in a permalloy film of $2$~$\mu$m$~\times~2$~$\mu$m where the static field is directed along the $y$-axis and the $23$~GHz oscillatory field is applied along the $z$-axis over a $100$~nm$~\times~100$~nm excitation region in the centre of the flim. Periodic boundary conditions are assumed in these simulations.

\begin{figure}
\includegraphics[width=6.6in]{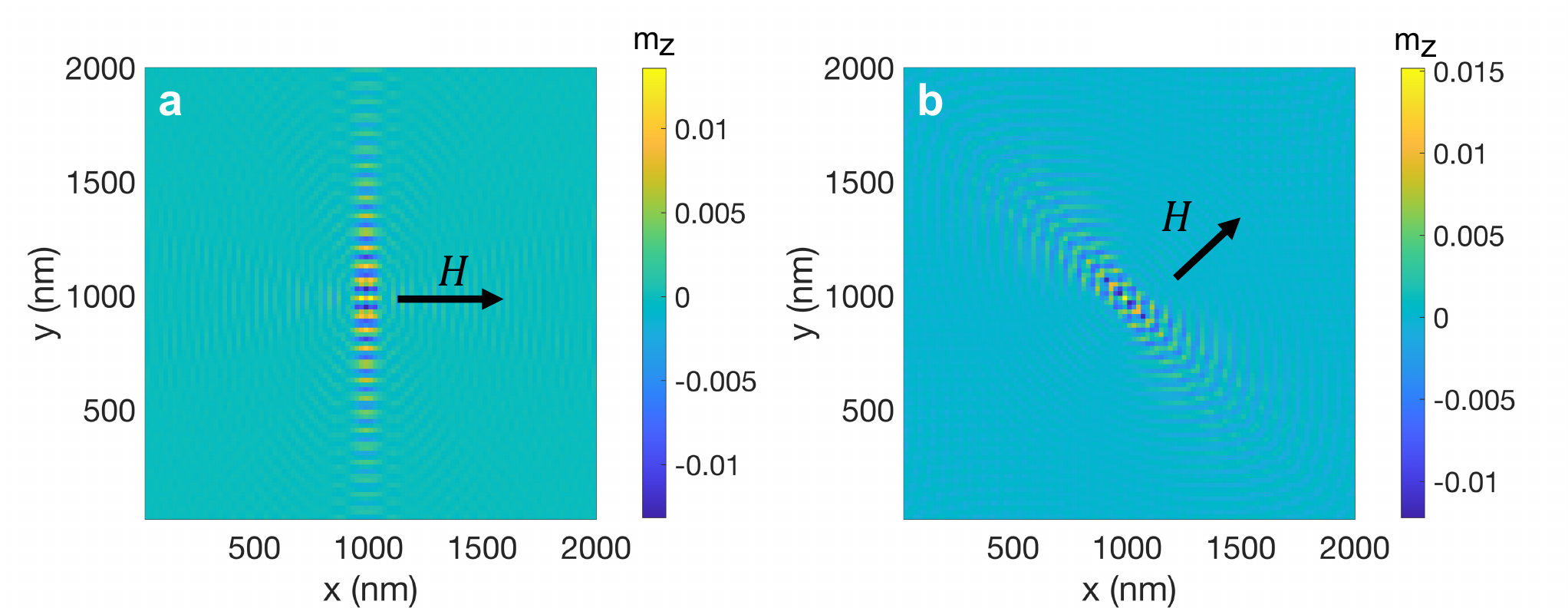}
\centering
\caption{\label{CausticBeam} \textbf{Caustic wave beams.} \textbf{a} Point source field directed in plane to sample with resulting caustic beam propagating perpendicular to field. $H$ displays the direction of the in-plane field. Colourbar displays amplitude of wave. \textbf{b} Point source field directed in plane to sample with resulting caustic beam propagating perpendicular to field. The direction of the in-plane field is displayed by an arrow. The colourbar displays amplitude of the wave.}
\end{figure}

The resulting caustic wave propagates normal to the static field orientation, as seen in Fig.~\ref{CausticBeam}\textbf{a}. To demonstrate that the dispersion relation can be arbitrarily rotated in plane, we change the angle of the static field by $45$~deg. The resulting caustic wave is displayed in Fig.~\ref{CausticBeam}\textbf{b}. In this case, the beam is not exactly collimated because the excitation is at an angle from a square oscillating region. However, it clearly propagates predominantly perpendicular to the external field.

These simulations demonstrate that the dipole-exchange kernel can be used for applications that require directional energy transport and for magnonic devices~\cite{Swyt2024}.

\subsection*{Scattering from defects}

A crucial effect to model is magnon scattering, particularly at physical boundaries and defects. This is important for magnonic waveguides~\cite{Chumak2015} as well as recent inverse design approaches~\cite{Papp2021,Wang2021}. As a proof of concept, we performed simulations of a $4$~$\mu$m$~\times~4$~$\mu$m permalloy thin film with five $200$~nm$\times 200$~nm defects placed randomly throughout the film. The magnetization is initialized in a uniform state along the $y$-axis, parallel to the 79~kA~m$^{-1}$ applied field. The magnons are excited from the left edge by an oscillating field at $11$~GHz oriented along the $x$-axis and applied over a width of $200$~nm. We implement reflecting, or free-spin, boundary conditions in this case.

\begin{figure}
\includegraphics[width=\linewidth]{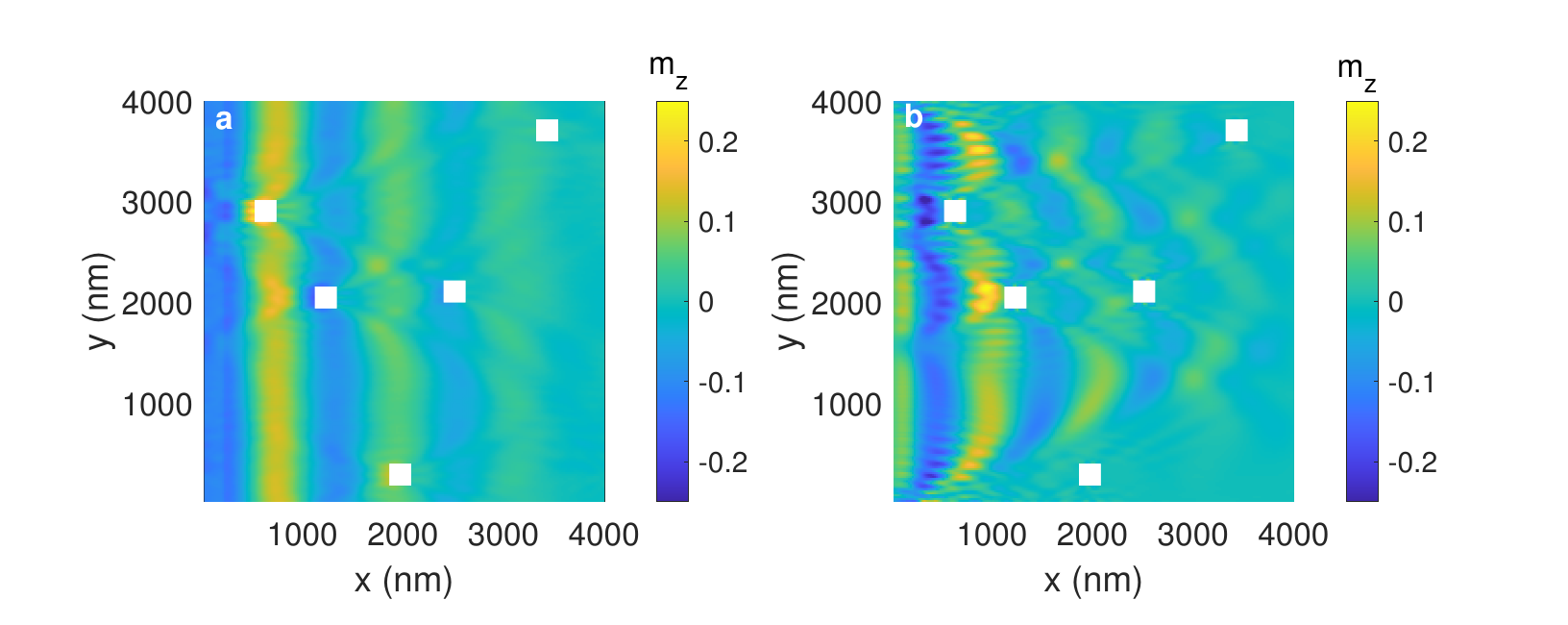}
\centering
\caption{\label{fig:defects} \textbf{Magnon scattering with defects}. Magnons are inserted from the left side of the thin film by use of an oscillatory field. The five white boxes represent defects as material voids. The spin waves interact with the defects causing wave scattering and interference. Snapshots of the $m_z$ magnetization component are shown for \textbf{a} the dipole-exchange kernel and \textbf{b} the magnon kernel with full nonlocal dipole calculation. Clearly, the full dipole affect the computation at the edges of both the defects and simulation domain. However, both approaches resolve magnon scattering.}
\end{figure}

A snapshot of the $m_z$ magnetisation component after $2$~ns of simulation time is shown in Fig.~\ref{fig:defects}\textbf{a} when using the dipole-exchange kernel. The local approximation of the demagnetising field precludes magnetisation bending at the edges of the simulated domain and defects. This also implies that the excited magnon is essentially a flat wavefront. Regardless, magnon scattering is observed due to the defects, evidenced by the V-shaped distortions in the waves behind the defects. These distortions overlap and lead to a characteristic diamond pattern in the centre of the simulation domain. For comparison, we performed the same simulation using the magnon kernel and full dipole computation. In this case, we first relax the simulation to account for the magnetisation edge bending and then excite a magnon with the oscillatory field. The corresponding snapshot of the $m_z$ magnetisation component at $2$~ns is shown in Fig.~\ref{fig:defects}\textbf{b}. Clearly, the magnon excitation is not flat but exhibits a characteristic triangular pattern that is due to the dipole field at the top and bottom edges of the simulated domain. These transverse components lead to additional magnon scattering events. Regardless of these additional effects, the same magnon scattering processes due to interactions with defects are observed, including the characteristic diamond-shape patterns in the centre of the simulation.

\section*{Conclusion}

We have demonstrated the use of a dipole-exchange kernel in the PS-LL formalism that allows for the accurate modelling of magnetostatic waves. Because the approach is spectral, the nonlocal interactions can be directly considered without the computation of the nonlocal dipole field, which requires at least six additional FFT computations per integration step. For this reason, our approach, compared to using the Newell tensor, increases computational speed by at least a factor two using a CPU.

It is important to emphasise the limitations of this model. First, a nonlocal dipole field is necessary to compute the correct stable magnetisation state of finite materials. This effect becomes increasingly important as the material thickness increases. Therefore, the dipole-exchange kernel is only valid for thin films and relatively extended films where physical boundary conditions have a negligible impact. Second, and related to the first point, is the requirement that the thickness of the material must be such that there is no magnon interaction between different thickness modes. This is the most stringent condition regarding the allowed thickness of the modeled materials. However, magnonic research is often conducted at frequencies where one magnon is efficiently excited and does not scatter with higher-order thickness modes or phonons. To consider these cases, it is necessary to define a three-dimensional kernel and simulation so that the thickness modes can be numerically resolved uniquely. Third, the simulation in itself is completely nonlinear, but the kernel is derived from linear waves. This means that nonlinear solutions, such as solitons, are not accurately resolved. Fourth, the model is, so far, only valid in-plane. It is possible to generalize it using the full formalism derived in Ref.~\cite{Kalinikos1986} but in this case, one would also need  to carefully consider the boundary conditions at the top and bottom surfaces of the thin film.

In summary, our approach is targeted at magnonic simulations relying on well-defined magnetostatic magnons excited in thin films. The acceleration achieved with our approach makes it attractive for machine-learning-based investigations, such as inverse design magnonics. The implementation of the dipole-exchange kernel also shows the versatility of the PS-LL formalism to resolve atomic scale dynamics on par with micrometre-scale dynamics.

\section*{Methods}

\subsection*{Magneto-dipolar kernel derivation}

The dispersion relation in the PS-LL model can be directly given in the kernel for symmetric geometries~\cite{Rockwell2024,Copus2025}. However, for asymmetric dispersion relations, such as that of magnetostatic waves, the kernel must be carefully derived.

Starting from the PS-LL model, Eqs.~\eqref{eq:LL} and \eqref{eq:Heff}, we can derive the dispersion relation of an in-plane uniformly magnetized thin film. For this, we set $\alpha=0$ so that only the conservative term of the dynamic equation is solved to compute the eigenvalues. The effective field is set to
\begin{equation}
    \label{eq:HeffSimple}
    \mathbf{h}_l = h_0\hat{x}-m_z\hat{z}.
\end{equation}

This field implies that the magnetization is saturated in the $\hat{x}$ direction. Thus, we consider $m_x\approx1$ and set the equations of motion for $m_y$ and $m_z$
\begin{subequations}
    \label{eq:PSLL_small}
    \begin{eqnarray}
        \label{eq:PSLL_my}
        \frac{\partial}{\partial t}m_y &=& -\gamma\mu_0M_sm_zh_0+m_z\mathcal{F}^{-1}\{\kappa(\mathbf{k})\hat{m_x}\}-\gamma\mu_0M_sm_z-\mathcal{F}^{-1}\{\kappa(\mathbf{k})\hat{m_z}\},\\
        \label{eq:PSLL_mz}
        \frac{\partial}{\partial t}m_z &=& -\mathcal{F}^{-1}\{\kappa(\mathbf{k})\hat{m_y}\}+\gamma\mu_0M_sm_yh_0-m_y\mathcal{F}^{-1}\{\kappa(\mathbf{k})\hat{m_x}\}.
    \end{eqnarray}
\end{subequations}

Because $m_x=1$, it follows that $\hat{m_x}=\delta(\mathbf{k})$. Furthermore, we enforce a kernel such that $\kappa(|\mathbf{k}|=0)=0$ so that $\kappa(\mathbf{k})\hat{m_x}\equiv0$. In addition, we can use the nomenclature of the main text where $\omega_H=\gamma\mu_0M_sh_0$ and $\omega_M=\gamma\mu_0M_s$ to rewrite the equations as
\begin{subequations}
    \label{eq:PSLL_small2}
    \begin{eqnarray}
        \label{eq:PSLL_my2}
        \frac{\partial}{\partial t}m_y &=& -\omega_Hm_z-\omega_Mm_z-\mathcal{F}^{-1}\{\kappa(\mathbf{k})\hat{m_z}\}\\
        \label{eq:PSLL_mz2}
        \frac{\partial}{\partial t}m_z &=& \mathcal{F}^{-1}\{\kappa(\mathbf{k})\hat{m_y}\}+\omega_Hm_y.
    \end{eqnarray}
\end{subequations}

We now perform a Fourier transform so that $\partial/\partial t-\rightarrow-i\omega$ and write the equations in matrix form
\begin{equation}
    \label{eq:PSLL_matrix}
    \begin{bmatrix}-i\omega&\omega_H+\omega_M+\kappa(\mathbf{k})\\\omega_H+\kappa(\mathbf{k})&i\omega\end{bmatrix}\begin{bmatrix}\hat{m_y}\\\hat{m_z}\end{bmatrix}=0
\end{equation}

Solving this eigenvalue problem leads to Eq.~\eqref{eq:omega_nl} which reduces to Kittel's equation when $|\mathbf{k}|=0$, as expected from our kernel definition.

\subsection*{Dipole kernel and boundary conditions}

The full dipole calculation is implemented in the PS-LL model using the Newell tensor approach~\cite{Newell1993}. For this, the space is extended in all directions. Even though we use a 2D simulation, this implies that a second cell is considered along the $\hat{z}$, normal-to-plane direction. The implementation is thus naturally three-dimensional. Using a single cell along the thickness simply leads to one layer with meaningful information.

The Newell tensor requires doubling the size of the computational domain, which has implications in how our boundary conditions are implemented. Notably, we distinguish between the boundaries used for the dipole calculation and those used for the wave boundaries. In the case of the dipole, periodic boundary conditions are specified by considering half a repetition in the $\hat{x}$ and $\hat{y}$ spatial orientations, including the corners. The $\hat{z}$ orientation is considered to be a free spin because we are focusing on thin films. Free spin conditions are simply specified by setting a zero magnetisation in the extended domain. 

For wave boundaries, the periodic boundary conditions are identical to those used for the dipole calculation. However, for free spin boundary conditions, wave reflection is achieved with Neumann boundary conditions. This means that the magnetisation is considered to be uniform and equal to the edge magnetisation in an extended domain. Furthermore, a two-dimensional Hann window is used in the extended domain to eliminate wave reflections. This ensures that the edges of the extended domain are equal to zero so that the Fourier transform is meaningful. Note that a window function is not needed for the dipole calculation because the Newell tensor is composed of decaying functions.

\subsection*{PS-LL implementation}

To implement the dipole-exchange dispersion relation within our simulations, we numerically construct a dipole-exchange kernel in Fourier space. We evaluate Eq.~\ref{eq:omega_nl} over each ($k_x, k_y$) cell using the magnon dispersion relation (Eq.~\ref{eq:omega2D}) for the exchange interaction. The angular terms for $F(\mathbf{k})$ in Eq.~\ref{eq"Fnn} are determined by rotating the wavevector direction $\varphi = \tan^{-1} (k_y/k_x)$ according to the direction of the external field $\varphi \rightarrow \varphi~-~\tan^{-1} (H_y/H_x)$. Once $\omega^2_\mathrm{DE}$ is obtained for each ($k_x, k_y$), the dipole-exchange kernel for the simulation is computed with Eq.~\ref{eq:kernel}.

\section*{Acknowledgments}

This work was supported by the U.S. Department of Energy, Office of Basic Energy Sciences under Award No. DE-SC0024339.

\section*{Author Contributions}

A.R. and E.I. derived the dipole-exchange kernel. A.R. and M.C. implemented the dipole-exchange kernel in the 2D PS-LL model. All authors contributed with performing validation simulations, analyzing the data, and writing the paper.

\section*{Competing interests statement}

All authors declare no financial or non-financial competing interests. 

\section*{Corresponding authors}
\noindent Correspondence to\\eiacocca@uccs.edu

\end{document}